\documentclass[11pt]{article}
\usepackage{natbib,moriond}
\usepackage{amsmath}
\usepackage{amssymb}
\usepackage{xspace}

\usepackage{epstopdf}
\usepackage{thumbpdf}

\def\Journal#1#2#3#4{{#1} {\bf #2}, #3 (#4)}

\newcommand{\rmn}{\mathrm}
\def\QLF{\phi_Q}
\def\BLF{\phi_B}
\def\Fermi{{\em Fermi}\xspace}

\newcommand{\Photo}{}

\bibpunct{(}{)}{;}{a}{,}{,}

\begin{document}
\vspace*{4cm}
\title{THE PHYSICS AND COSMOLOGY OF TEV BLAZARS IN A NUTSHELL}

\author{ C.~PFROMMER$^1$, 
A.~E.~BRODERICK$^{2}$,
P.~CHANG$^3$,
E.~PUCHWEIN$^1$,
V.~SPRINGEL$^{1,4}$ }

\address{
  ${}^{1}$ {Heidelberg Institute for Theoretical Studies, Schloss-Wolfsbrunnenweg 35, 69118 Heidelberg, Germany}\\
  ${}^{2}$ {Perimeter Institute for Theoretical Physics, 31 Caroline Street North, Waterloo, ON, N2L 2Y5, Canada;
   Department of Physics and Astronomy, University of Waterloo, 200 University Avenue West, Waterloo, ON, N2L 3G1, Canada}\\
  ${}^{3}$ {Department of Physics, University of Wisconsin-Milwaukee, 1900 E. Kenwood Boulevard, Milwaukee, WI 53211, USA}\\
  ${}^{4}$ {Zentrum f\"ur Astronomie der Universit\"at Heidelberg,
    Astronomisches Recheninstitut, M\"onchhofstr. 12-14, 69120
    Heidelberg, Germany}}

\maketitle\abstracts{The extragalactic gamma-ray sky at TeV energies is
  dominated by blazars, a subclass of accreting super-massive black holes with
  powerful relativistic outflows directed at us. Only constituting a small
  fraction of the total power output of black holes, blazars were thought to
  have a minor impact on the universe at best. As we argue here, the opposite is
  true and the gamma-ray emission from TeV blazars can be thermalized via
  beam-plasma instabilities on cosmological scales with order unity efficiency,
  resulting in a potentially dramatic heating of the low-density intergalactic
  medium. Here, we review this novel heating mechanism and explore the
  consequences for the formation of structure in the universe. In particular, we
  show how it produces an inverted temperature-density relation of the
  intergalactic medium that is in agreement with observations of the
  Lyman-$\alpha$ forest. This suggests that {\it blazar heating} can potentially
  explain the paucity of dwarf galaxies in galactic halos and voids, and the
  bimodality of galaxy clusters. This also transforms our understanding of the
  evolution of blazars, their contribution to the extra-galactic gamma-ray
  background, and how their individual spectra can be used in constraining
  intergalactic magnetic fields.}

\section{Introduction}
\label{sec:intro}

The extragalactic gamma-ray sky is dominated by ``blazars''. These are a
subclass of super-massive black holes, situated at the center of every galaxy,
which drive powerful relativistic jets and electromagnetic radiation out to
cosmological distances. An important subset of blazars exhibit hard power-law
spectra that extend to TeV photon energies (high-energy-peaked BL Lacs). The
Universe is opaque to the emitted TeV gamma rays because they annihilate and
pair produce on the extragalactic background light which is emitted by galaxies
and quasars through the history of the universe. The mean free path for this
reaction is $\lambda_{\gamma\gamma} \sim (700\ldots35)\, (E/\rmn{TeV})^{-1}$~Mpc
for redshifts $z=0\ldots1$, respectively, and is approximately constant at
$\lambda_{\gamma\gamma} \sim 35\, (E/\rmn{TeV})^{-1}$~Mpc for higher
redshifts. The resulting ultra-relativistic pairs of electrons and positrons are
commonly assumed to lose energy primarily through inverse Compton scattering
with photons of the cosmic microwave background, cascading the original TeV
emission a factor of $\sim10^3$ down to GeV energies.

However, there are two serious problems with this picture: the expected cascaded
GeV emission is not seen in the individual spectra of those blazars (Neronov \&
Vovk 2010) and the emission of all unresolved blazars would overproduce the
observed extragalactic gamma-ray background (EGRB) at GeV energies {\it if}
these objects share a similar cosmological evolution as the underlying black
hole or parent galaxy population (Venters 2010). As a putative solution to the
first problem, comparably large magnetic fields have been hypothesized which
would deflect the pairs out of our line-of-sight to these blazars (Neronov \&
Vovk 2010), diluting the point-source flux into a lower surface brightness
``pair halo''.  However, magnetic deflection of pairs (and hence their inverse
Compton emission) out of our line-of-sight is on average balanced by deflecting
other pairs into our line-of-sight, so that the resulting isotropic EGRB remains
invariant. This represents a substantial problem to unifying the hard gamma-ray
blazar population with that of other active galactic nuclei (AGN), is at odds
with the underlying physical picture of accreting black hole systems, and
suggests an unlikely conspiracy between accretion physics and the formation of
structure.

\section{Beam-plasma instabilities}

\begin{figure}
\begin{minipage}{0.32\linewidth}
\centerline{\includegraphics[width=\linewidth]{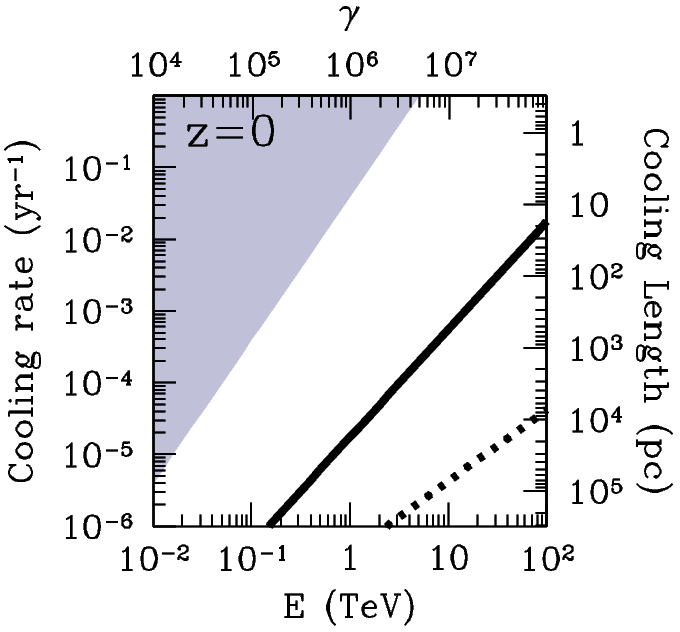}}
\end{minipage}
\hfill
\begin{minipage}{0.32\linewidth}
\centerline{\includegraphics[width=\linewidth]{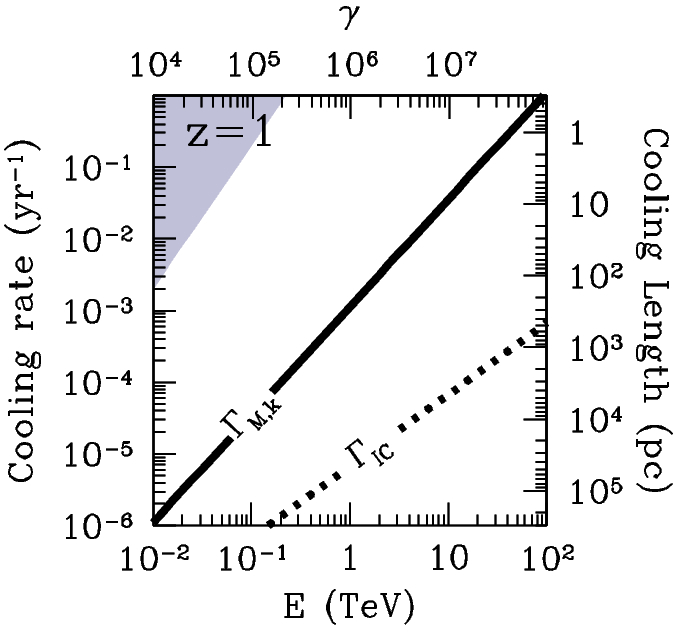}}
\end{minipage}
\hfill
\begin{minipage}{0.32\linewidth}
\centerline{\includegraphics[width=\linewidth]{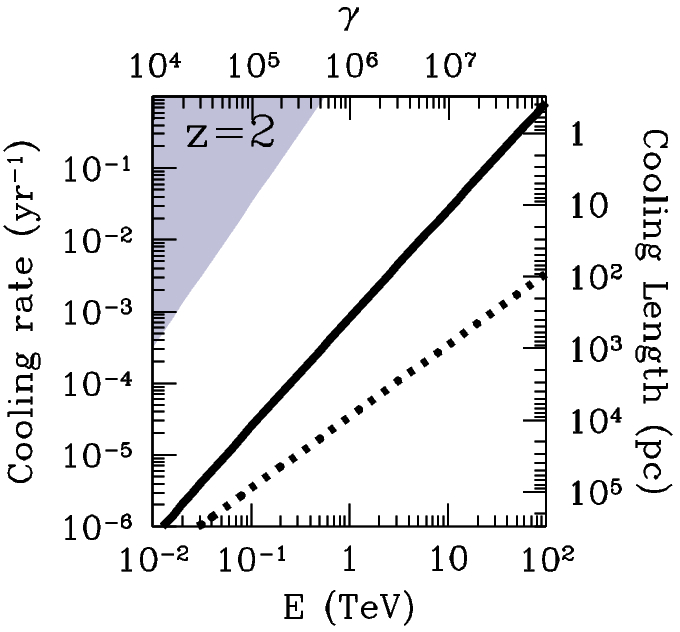}}
\end{minipage}
\caption{Initial pair beam cooling rates due to the kinetic oblique instability
  (thick solid) and inverse Compton scattering (dotted) as a function of
  gamma-ray energy ($E$) at a number of redshifts ($z$).  In all cases, we
  consider a mean-density region, and the isotropic-equivalent luminosity of the
  source at energy $E$, $E L_E$, is $10^{45}\,\rmn{erg}\,\rmn{s}^{-1}$, similar
  to the brightest TeV blazars seen from Earth. We list the initial pair Lorentz
  factor, $\gamma$, and cooling lengthscale along the top and right axes,
  respectively (from Broderick {\it et al.}  2012).}
\label{fig:Gamma}
\end{figure}

Recently, we have shown that there is an even more efficient mechanism that
competes with this cascading process. Plasma instabilities driven by the highly
anisotropic nature of the ultra-relativistic pair distribution provide a
plausible way to dissipate the kinetic energy of the TeV pairs locally, heating
the intergalactic medium (Broderick {\it et al.} 2012). We can understand the
two-stream instability intuitively by considering a longitudinal wave-like
perturbation of the charge of the background plasma along the beam direction
(i.e., a Langmuir wave). The initially homogeneous beam electrons feel repulsive
(attractive) forces by the potential minima (maxima) of the electrostatic wave
in the background plasma. As a result, electrons (positrons) attain their lowest
velocity in the potential minima (maxima), which causes them to bunch up. Hence,
the bunching within the beam is simply an excitation of a beam Langmuir wave
that couples in phase with the background perturbation. This enhances the
background potential and implies stronger forces on the beam pairs. This
positive feedback loop causes exponential wave-growth, i.e.{\ }the onset of an
instability. In practice, oscillatory modes that propagate in an oblique
direction to the beam grow substantially faster than the two-stream instability
just discussed. The reason is that electric fields can more easily deflect
ultra-relativistic particles than change their parallel velocities (see
Broderick {\it et al.} 2012, for details).

Unstable electromagnetic waves grow fastest when the velocity dispersions are
smallest across their wave fronts. As these velocity dispersions get larger and
larger, i.e., for increasing temperature, the growth rate of the unstable
oblique mode moves into the finite temperature or kinetic regime, where the
exponential growth rate is reduced due to the effects of phase mixing and
decoherence.  In Fig.~\ref{fig:Gamma}, we show the pair beam cooling rates due
to the kinetic oblique instability in the linear regime, $\Gamma_\rmn{M,k}$
(Bret {\it et al.} 2010a) for a beam density that obeys the steady-state
Boltzmann equation, i.e., we account for production and various loss processes
of the pairs. Most importantly, we find that $\Gamma_\rmn{M,k}$ dominates over
the inverse Compton cooling rate $\Gamma_\rmn{IC}$ by more than an order of
magnitude for the parameters of luminous TeV blazars.

Analytical quasi-linear calculations of the cold regime (Schlickeiser {\it et
  al.} 2012) and numerical work of the oblique instability in the kinetic regime
(Bret {\it et al.}  2010b) with smaller density contrasts than considered here
suggest that the dominance of the oblique instability carries over in the regime
of non-linear saturation, although there is currently a debate about the role of
induced scattering by thermal ions on this non-linear saturation (Miniati \&
Elyiv 2013, Schlickeiser {\it et al.} 2013, Chang {\it et al.} in prep.). In the
following, we assume that a large fraction of the free kinetic energy of the
pairs is transferred to the electromagnetic modes in the background plasma,
which should eventually be dissipated, heating the intergalactic medium (IGM).

\section{Implications for the blazar luminosity function and the gamma-ray sky}

\begin{figure}
\begin{center}
\includegraphics[width=0.49\linewidth]{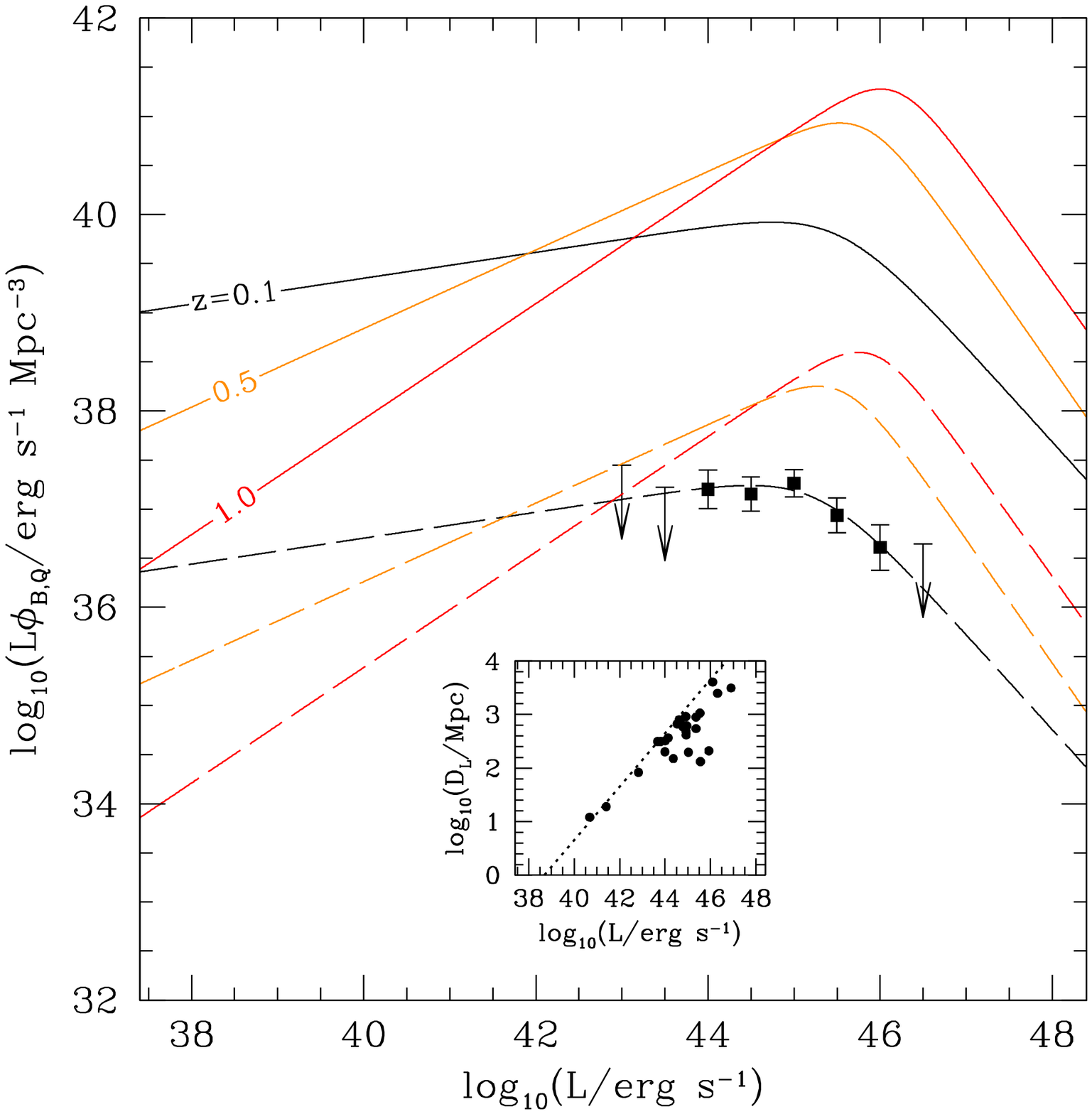}
\includegraphics[width=0.49\linewidth]{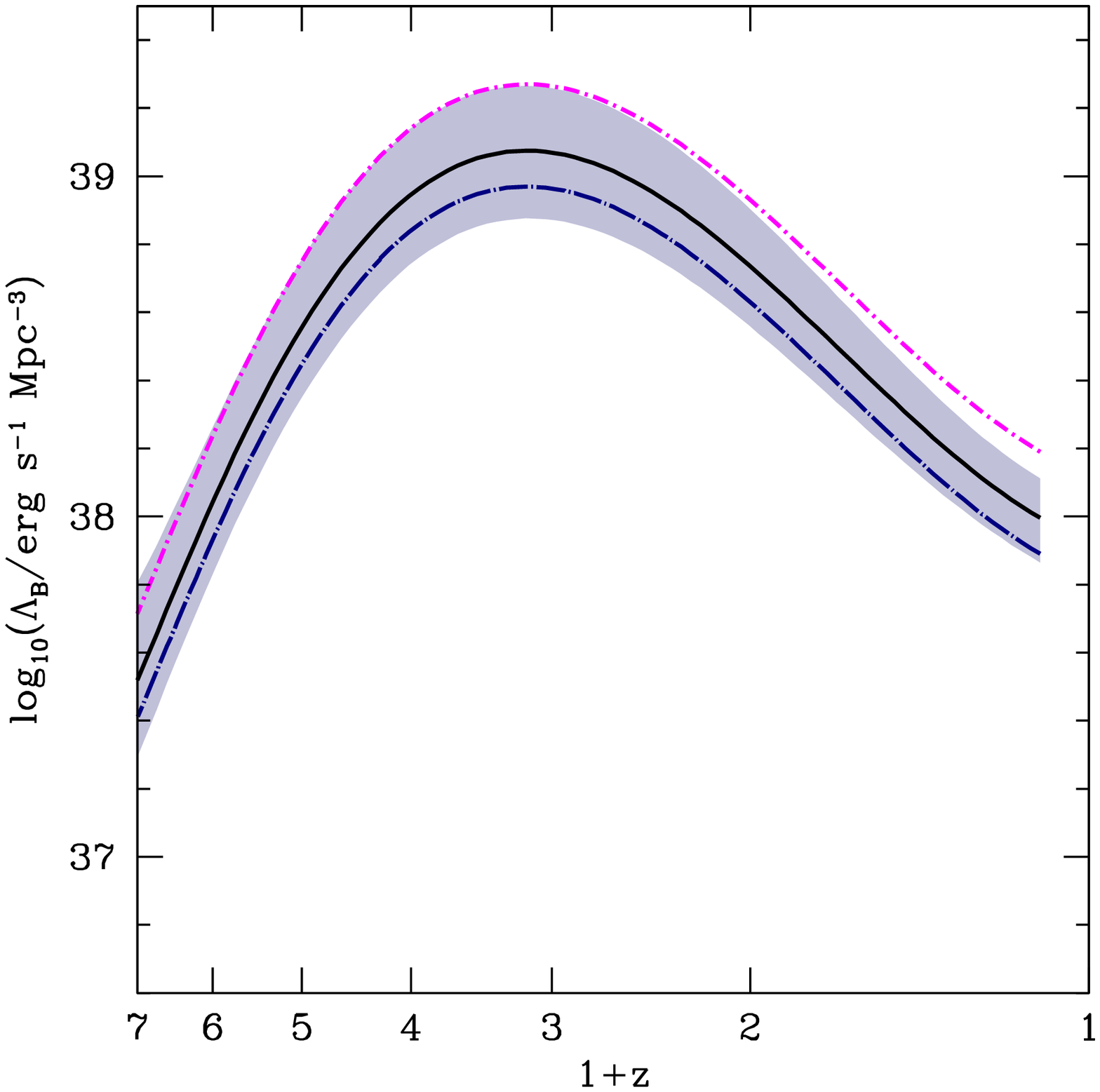}
\end{center}
\caption{{\it Left.} Comparison between the luminosity-weighted quasar and
  TeV-blazar luminosity functions ($L\QLF(z,L)$ and $L\BLF(z,L)$, respectively).
  The solid lines show the absolute $L\QLF$ (in comoving Mpc), while the dashed
  lines show $L\QLF$ rescaled in magnitude by $2.1\times10^{-3}$ and shifted to
  lower luminosities by a factor of $0.55$. Different redshifts are color coded
  as indicated in the figure. The points and upper-limits show $\BLF$ of all
  high- and intermediate-energy-peaked blazars with good spectral measurements.
  Presented in the inset is the TeV source luminosity distance as a function of
  source luminosity for all of the blazars with redshift estimates (including
  limits). The dotted line shows the distance-dependence of the flux limit we
  employ in the completeness correction (from Broderick {\it et al.}
  2012). {\it Right.} Comoving blazar luminosity density $\Lambda_B(z) =
  \int_{L_\rmn{min}}^{\infty} \!\!d L\, \BLF(z,L)$ as a function of redshift. The
  shaded region represents the 1-$\sigma$ uncertainty that results from a
  combination of the uncertainty in the number of bright blazars that contribute
  to the local heating and in the uncertainties in the quasar luminosity density
  (Hopkins {\it et al.} 2007) to which we normalize (from Chang {\it et al.}
  2012).}
\label{fig:BLF}
\end{figure}

To assess implications for the gamma-ray sky and the thermal evolution of the IGM, we
construct a blazar luminosity function (BLF). In Broderick {\it et al.} (2012),
we collect the luminosity of all 23 TeV blazars with good spectral measurements and
account for selection effects (sky coverage, duty cycle, galactic
occultation, TeV flux limit). The resulting BLF is shown in
Fig.~\ref{fig:BLF}. Most notably, the TeV blazar luminosity density is a scaled
version of that of quasars. This implies that quasars and TeV blazars appear to
be regulated by the same mechanism and are contemporaneous elements of a single
AGN population, i.e., the TeV-blazar activity does not lag quasar
activity. Hence we adopt the plausible assumption that both distributions trace
each other for all redshifts and work out the implications of this assertion.

\begin{figure}
\begin{center}
\includegraphics[width=0.49\linewidth]{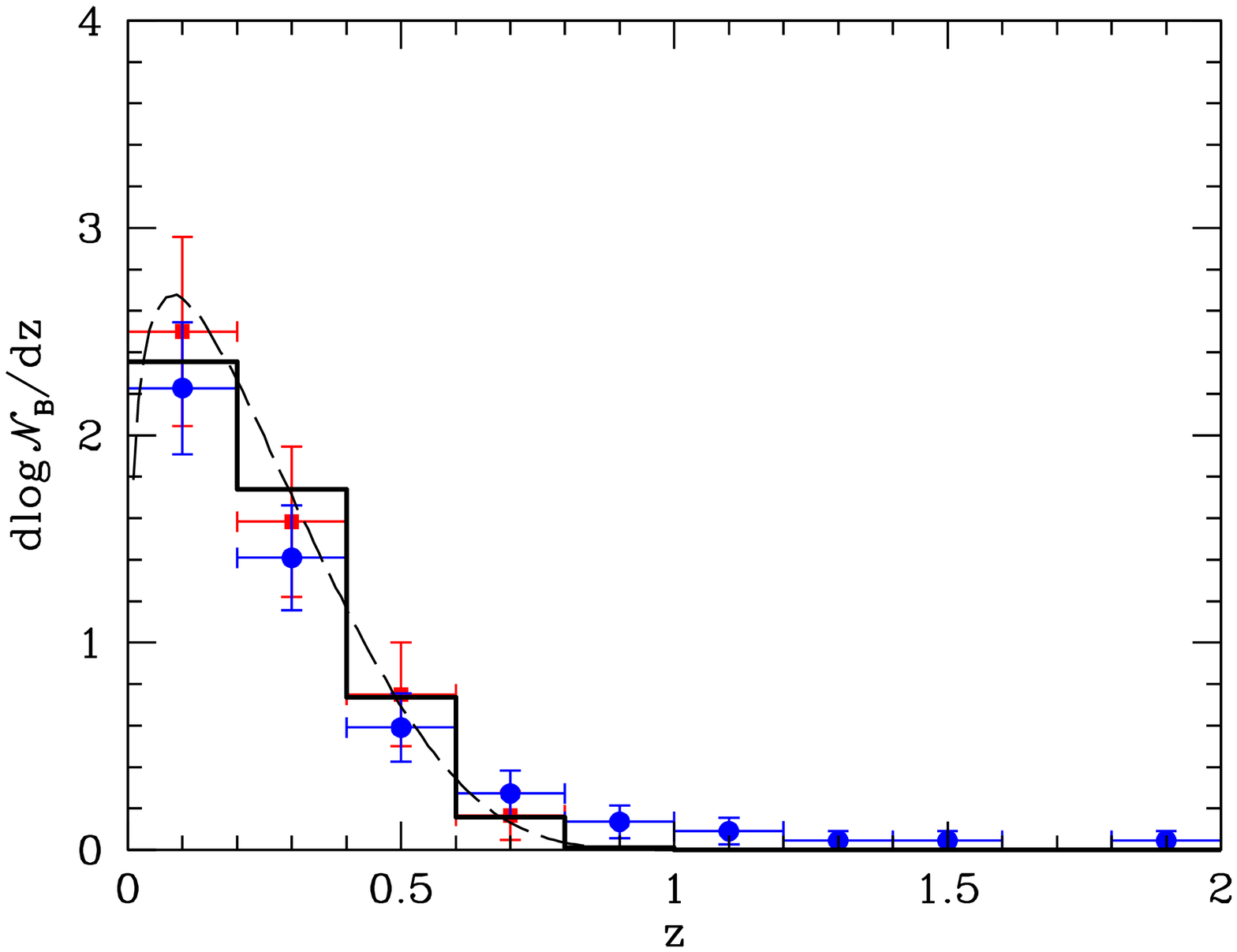}
\includegraphics[width=0.49\linewidth]{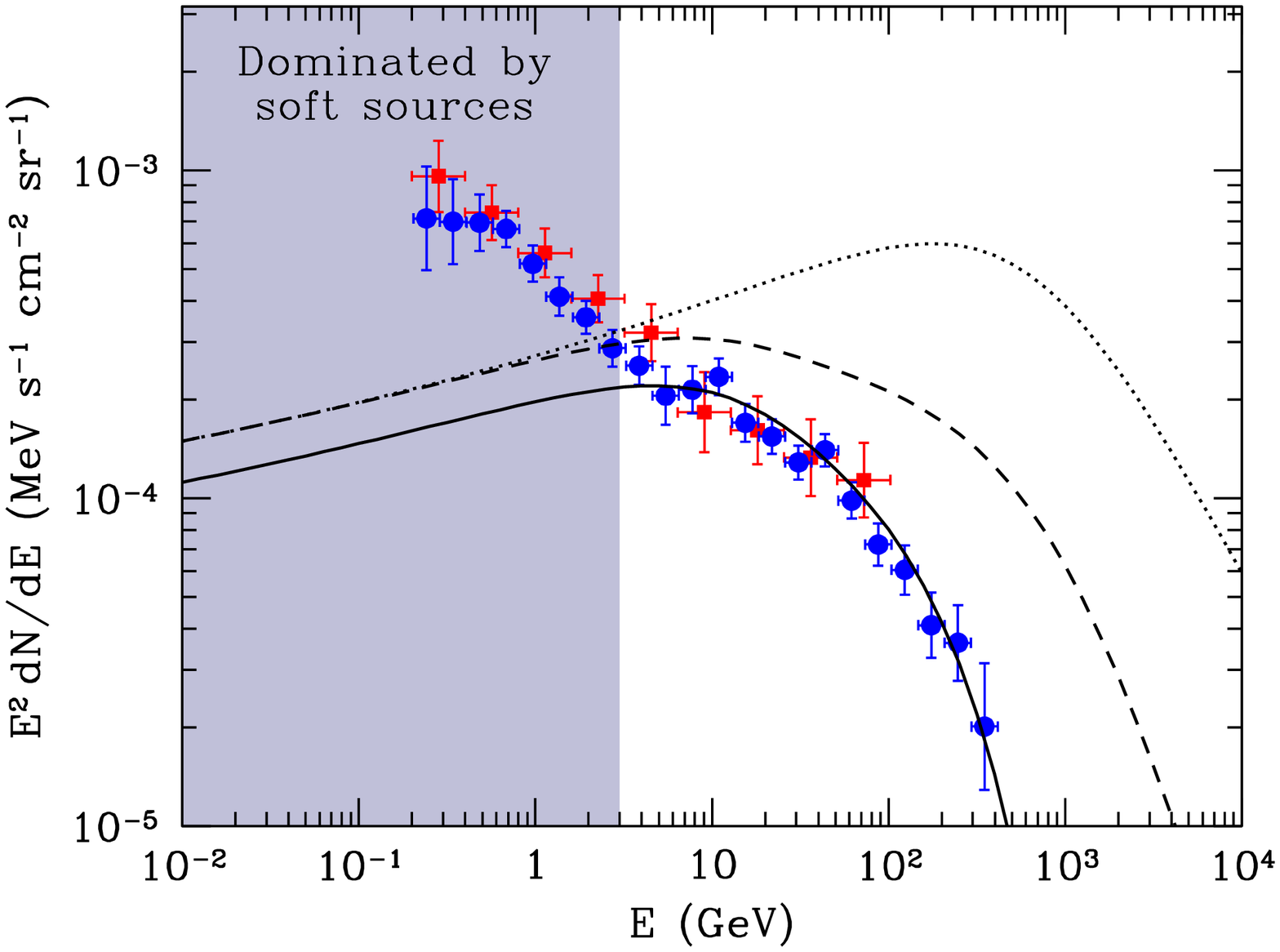}
\end{center}
\caption{{\it Left.} Nearby redshift distribution of the hard gamma-ray blazars
  above the \Fermi~flux limit, both in continuous form (dashed) and binned with
  $\Delta z=0.2$ (continuous). For comparison the redshift distribution of the
  \Fermi~hard gamma-ray blazars in the 1LAC (red squares) and 2LAC (blue
  circles) are also shown.  For these, the vertical error bars denote Poisson
  errors. {\it Right.}  \Fermi~EGRB anticipated by the hard gamma-ray blazars.
  The dotted, dashed, and solid lines correspond to the unabsorbed spectrum,
  spectrum corrected for absorption on the extragalactic background light, and
  spectrum additionally corrected for resolved point sources (assuming all hard
  gamma-ray blazars with $z\lesssim0.29$ are resolved).  These are compared with
  the measured \Fermi~EGRB reported in Abdo {\it et al.} (2010, red squares) and
  Ackermann {\it et al.} (2012, blue circles).  Note that below $\sim3$ GeV the
  EGRB is dominated by soft sources (from Broderick {\it et al.}  2013).}
\label{fig:EGRB}
\end{figure}

To quantify the impact on the gamma-ray sky, we need to expand the BLF to
include the intrinsic energy spectra, $dN/dE$, of blazars and adopt a typical
broken power-law spectrum
\begin{equation}
  \label{eq:dNdE}
  \frac{dN}{dE} = f \hat{F}_E =
    f \left[\left(\frac{E}{E_b}\right)^{\Gamma_l} + 
      \left(\frac{E}{E_b}\right)^{\Gamma_h}\right]^{-1},
\end{equation}
where $E_b\simeq1$~TeV is the break energy, $\Gamma_h \simeq 3$ is the high-energy
spectral index, and the intrinsic low-energy slope $\Gamma_l$ is softened with
increasing propagation length due to the higher probability of high-energy
photons to annihilate on the extragalactic background light. This yields a
steeper (larger) observed $\Gamma_F$, which we draw from the distribution of
local blazars as observed by the \Fermi gamma-ray telescope (that are not
affected by spectral softening due to pair production effects). We arrive at the
BLF, $d^4 \mathcal{N} / (d \log L_\rmn{TeV}\, dz\, dE\, d \Gamma_l)$, i.e., the
distribution of blazars with TeV luminosity $L_\rmn{TeV}$, redshift $z$,
gamma-ray energy $E$, and $\Gamma_l$.

Different projections of this BLF onto its independent variables allow
comparison to \Fermi data. Integrating this distribution over $L_\rmn{TeV}$, $E$
and $\Gamma_l$ and adopting integration limits that account for the \Fermi flux
limit $S_\rmn{min}$ yields the redshift distribution of \Fermi blazars (left
panel of Fig.~\ref{fig:EGRB}). Interestingly, an evolving (increasing) blazar
population is consistent with the observed declining number evolution of blazars
due to the \Fermi flux limit and the low intrinsic luminosity of the hard
blazars.  Masking these resolved blazars and integrating the blazar distribution
over $L_\rmn{TeV}$, $z$, and $\Gamma_l$ yields the contribution of blazars to
the {\it isotropic} EGRB (right panel of Fig.~\ref{fig:EGRB}).  This
demonstrates that an evolving population of hard blazars matches the latest data
of the EGRB by the \Fermi Collaboration at energies $\gtrsim 3$ GeV extremely
well. Moreover, the modeled $\log \mathcal{N}$-$\log S$ distribution and the
{\it anisotropic} EGRB, which mainly probes nearby objects below the
detectability limit, provide an excellent match to the \Fermi data (Broderick
{\it et al.} 2013). Hence, this naturally solves the two mysteries introduced in
Sect.~\ref{sec:intro} in a {\it unified model of blazars and their underlying
  black hole population without the need to invoke large magnetic
  fields}. Critical to this success is the absence of inverse Compton cascades
that would otherwise redistribute energy between the unabsorbed and the absorbed
spectrum into the energy range around 10 GeV, thus vastly overproducing the
tight limits provided by \Fermi.

\section{Rewriting the thermal history of the IGM and the Lyman-$\alpha$ forest}

\begin{figure}
\begin{center}
\includegraphics[width=0.49\linewidth]{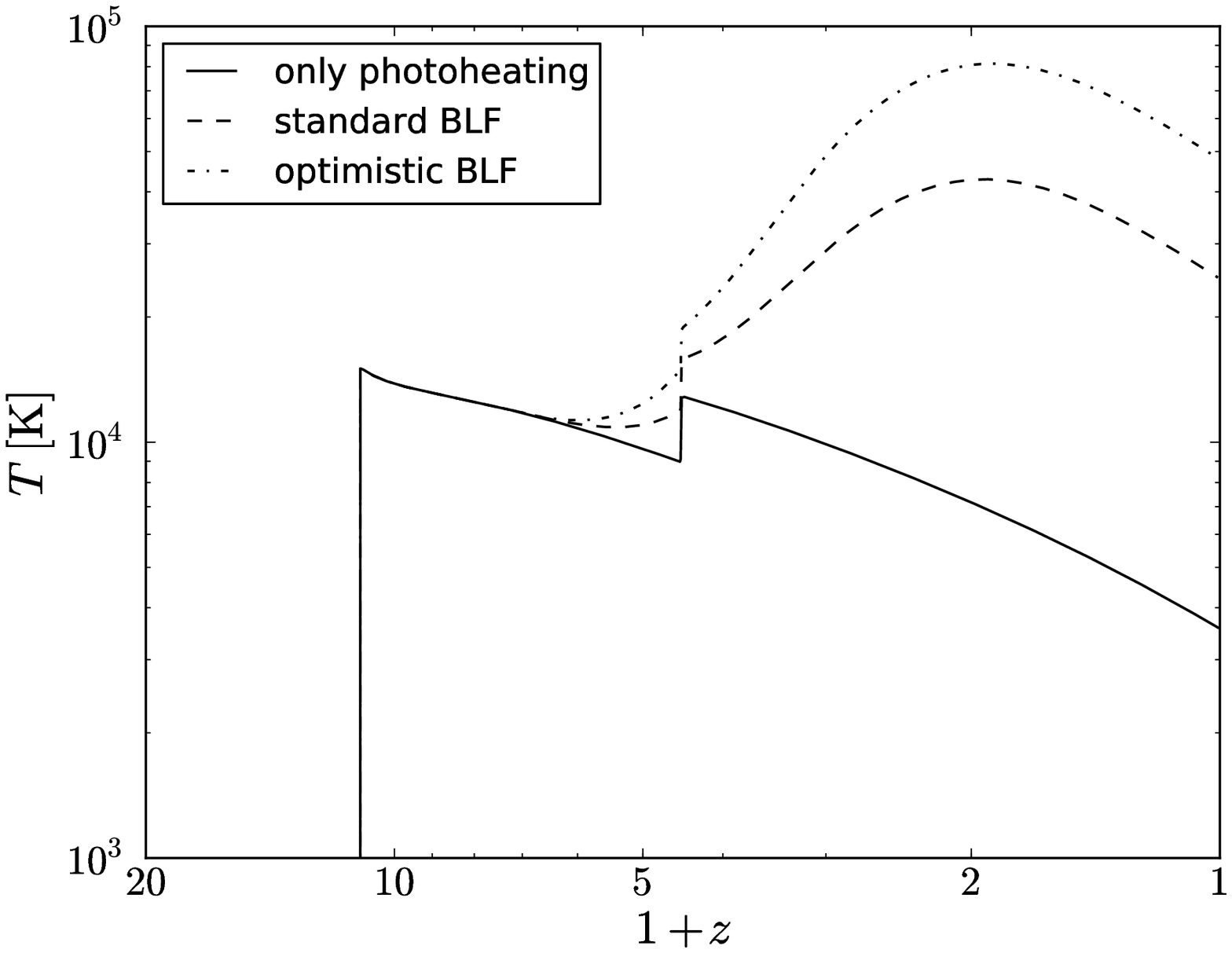}
\includegraphics[width=0.49\linewidth]{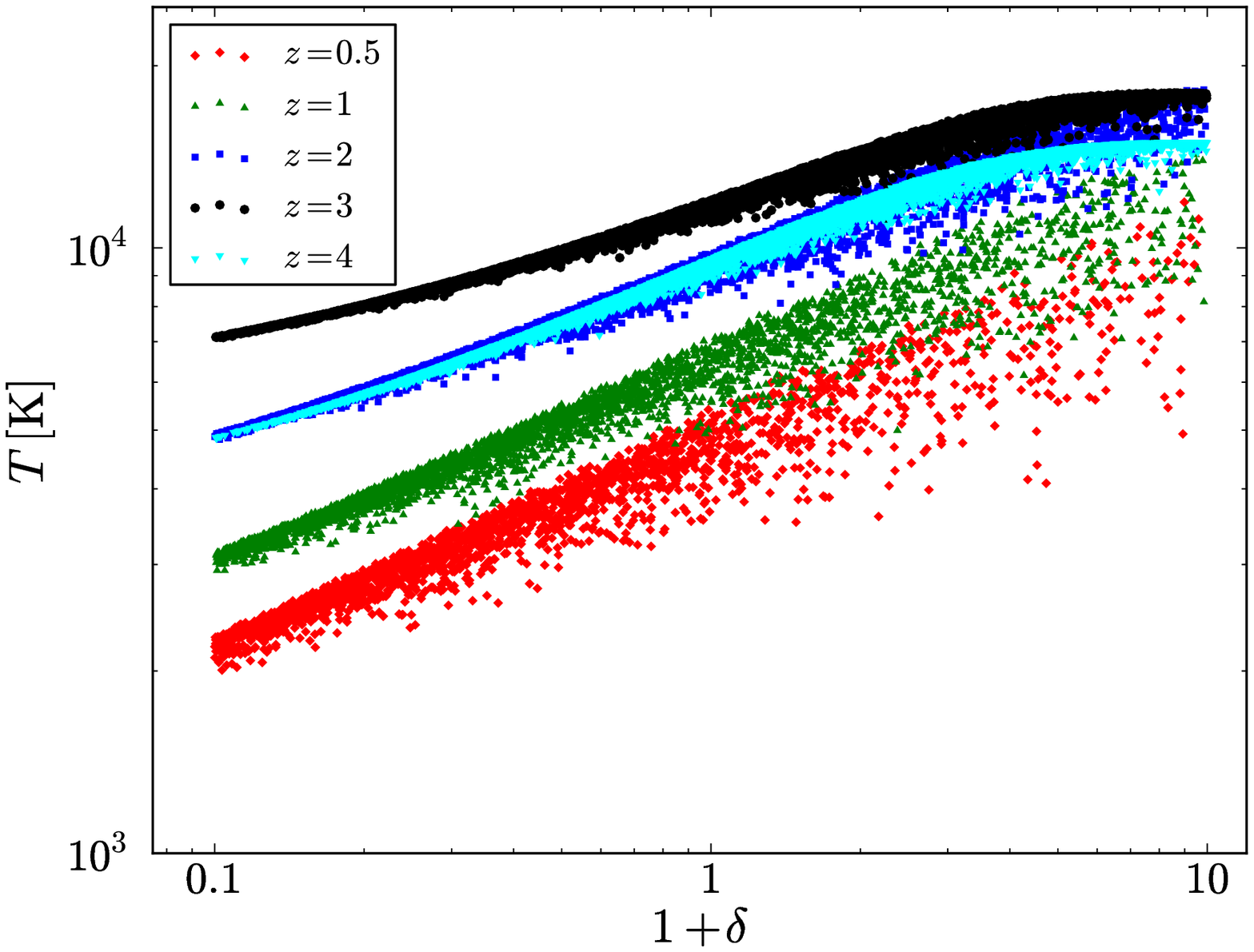}\\
\includegraphics[width=0.49\linewidth]{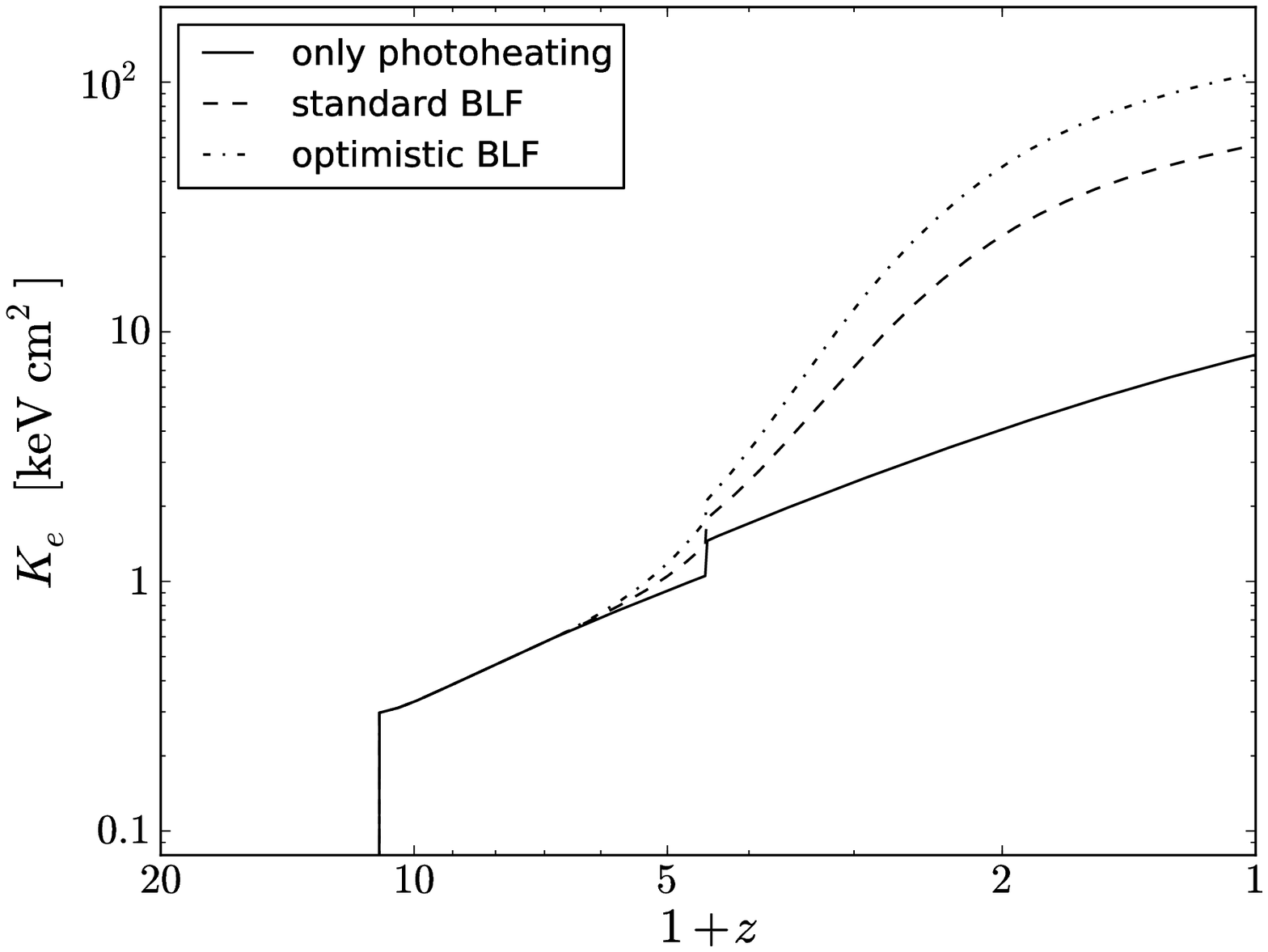}
\includegraphics[width=0.49\linewidth]{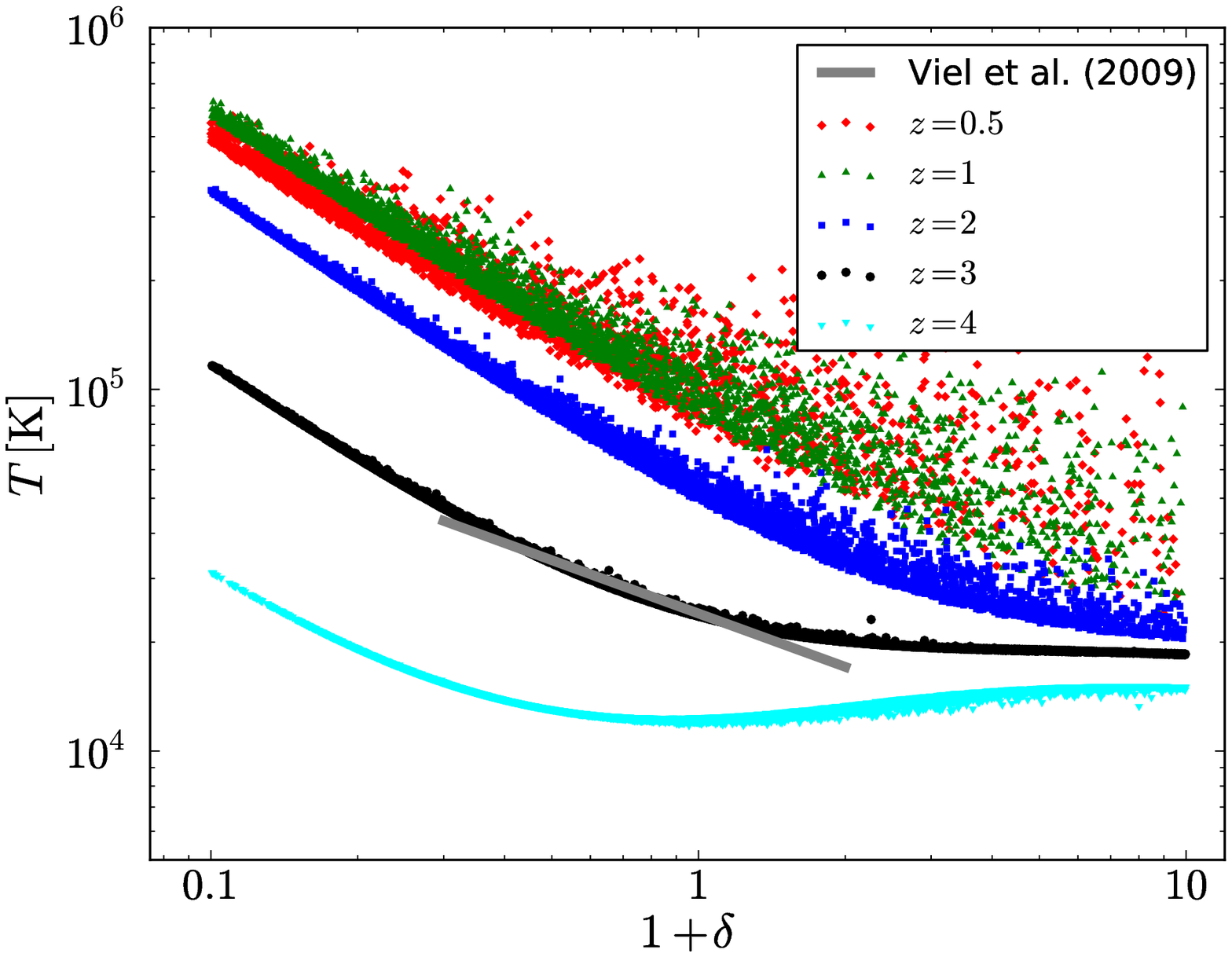}\\
\end{center}
\caption{Thermal history {\it (top left)} and entropy history {\it (bottom
    left)} of a patch of mean density ($\delta = 0$) of the IGM.  The solid
  curves are for pure photoheating with sudden reionization histories for H and
  He {\sc ii} at $z_{\rm reion} = 10$ and $z_{\rmn{He}\,\textsc{ii}} = 3.5$. The
  dashed (dash-dotted) lines show the evolution for the standard (optimistic)
  blazar heating model that employs the blazar luminosity density, i.e., using
  the redshift evolution of the quasar luminosity density and are normalized to
  the local heating rate, which is subject to an uncertain incompleteness
  correction factor (from Pfrommer {\it et al.} 2012). Temperature-density
  relation without blazar heating {\it (top right)} and for the optimistic
  blazar heating model {\it (bottom right)} at a number of redshifts (from Chang
  {\it et al.} 2012). The grey line is the best-fit model derived from
  Lyman-$\alpha$ data (Viel {\it et al.} 2004).}
\label{fig:T_z}
\end{figure}

\begin{figure}
\begin{center}
\includegraphics[width=0.49\linewidth]{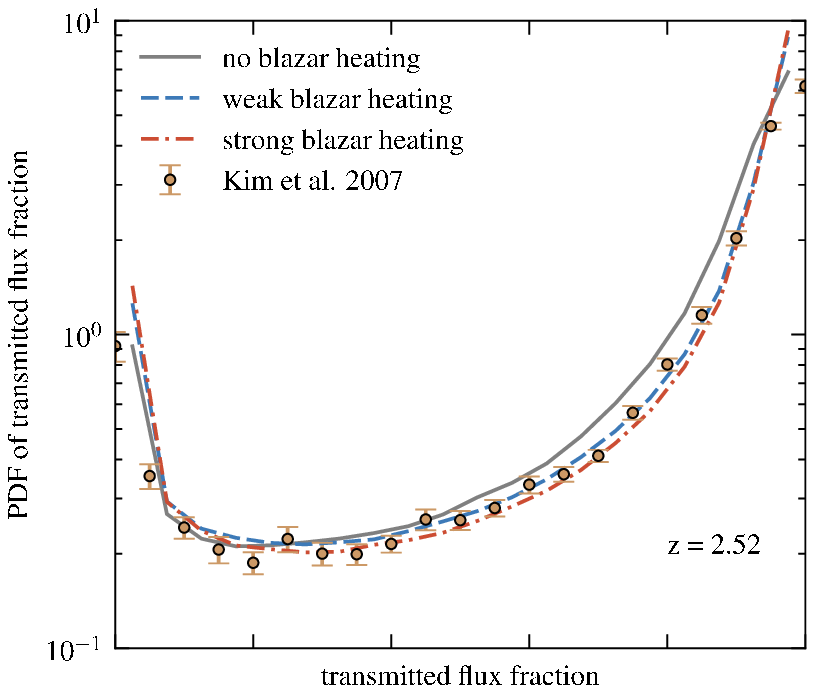}
\includegraphics[width=0.49\linewidth]{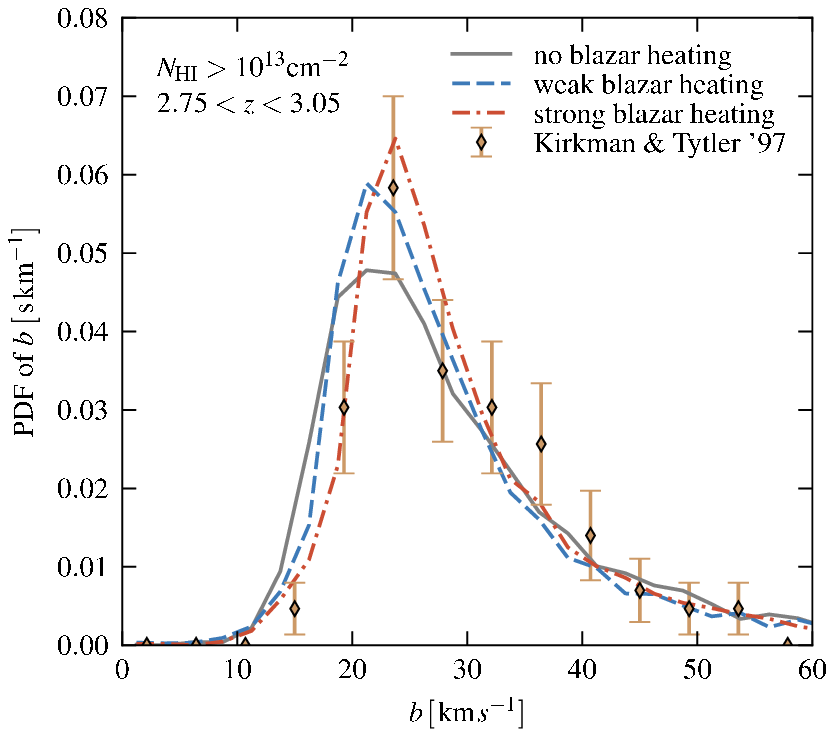}
\end{center}
\caption{Comparing the Lyman-$\alpha$ forest in hydrodynamical cosmological
  simulations with and without blazar heating.  {\it Left.} Probability
  distribution functions of the transmitted flux fraction for simulations with
  and without blazar heating for two different normalisations of the blazar
  heating rate are compared to observational constraints from Kim {\it et al.}
  2007.  {\it Right.} The normalised distribution function of Lyman-$\alpha$
  line widths, $b$, for simulations with and without blazar heating at redshift
  $z=3$ are compared to observational constraints by Kirkman {\it et al.}
  (1997). Both panels show simulation results for a UV background that was {\it
    matched} to the observed mean transmission (from Puchwein {\it et al.}
  2012).}
\label{fig:Lya}
\end{figure}

We find that for our BLF, every region in the universe is heated by at least one
TeV blazar back to $z\sim5$, providing a novel heating mechanism of the gas at
mean density that is ten times larger at the present time than what has been
previously considered (Chang {\it et al.} 2012).  This can be interpreted as a
gradually rising (and density dependent) entropy enhancement after $z=3$ (left
panels of Fig.~\ref{fig:T_z}). Unlike photoheating, the blazar heating rate per
unit volume does not depend on density since (1) the distributions of TeV
blazars and the extragalactic background light are uniform on the cosmological
scales of the mean free path of pair production, $\lambda_{\gamma\gamma}$, and
(2) it is nearly independent of the IGM density. Hence this particular heating
process deposits more energy per baryon in low-density regions and naturally
produces an inverted temperature-density relation in voids that reaches
asymptotically $T\propto 1/\rho$ (right panels of Fig.~\ref{fig:T_z}). This
unique property in combination with the recent and continuous nature of blazar
heating is needed to solve many problems present in previous calculations of
Lyman-$\alpha$ forest spectra.

Detailed cosmological simulations that include blazar heating show superb
agreement with all statistics used to characterize Lyman-$\alpha$ forest spectra
(Puchwein {\it et al.} 2012). In particular, our simulations with blazar heating
simultaneously reproduce the observed effective optical depth and temperature as
a function of redshift, the observed probability distribution functions of the
transmitted flux (Fig.~\ref{fig:Lya}), and the observed flux power spectra, over
the full redshift range $2<z<3$.  Additionally, by deblending the absorption
features of Lyman-$\alpha$ spectra into a sum of thermally broadened individual
lines, we find superb agreement with the observed lower cutoff of the line-width
distribution (Fig.~\ref{fig:Lya}) and abundances of neutral hydrogen column
densities per unit redshift.  This concordance between Lyman-$\alpha$ data and
simulation results, which are based on the most recent cosmological parameters,
also suggests that the inclusion of blazar heating alleviates previous tensions
on constraints of the normalization of the density power spectrum, $\sigma_8$,
derived from Lyman-$\alpha$ measurements and other cosmological data.

\section{Implications for the formation of dwarf galaxies and galaxy clusters}

We have seen that blazar heating dramatically changes the thermal history of the
diffuse IGM, which necessarily implies a number of important implications for
late-time structure formation (Pfrommer {\it et al.} 2012).  Unlike
photoionization models, which typically invoke the heating at reionization,
blazar heating provides a well defined, time-dependent entropy enhancement that
rises dramatically after $z\sim2$, suppressing the formation of late forming
dwarf galaxies. On small scales, thermal pressure opposes gravitational
collapse. This introduces a characteristic length and mass scale below which
galaxies do not form. A hotter intergalactic medium implies a higher thermal
pressure and a higher Jeans mass $M_J$ at redshift $z$,
\begin{equation}
  M_J \propto \frac{c_s^3(z)}{\sqrt{G^3\rho(z)}}
      \propto \left(\frac{T^3(z)}{G^3\rho(z)}\right)^{1/2}
  \quad\to \quad
  \frac{M_{J, \rmn{blazar}}}{M_{J, \rmn{photo}}} \approx
  \left(\frac{T_\rmn{blazar}}{T_\rmn{photo}}\right)^{3/2} \gtrsim 30,
\end{equation}
where $c_s$, $\rho$, and $T_\rmn{IGM}$ are the sound speed, density, and
temperature of the IGM, respectively, and $G$ is Newton's gravitational
constant. That is, blazar heating increases $M_J$ by 30 over pure photoheating
models.

\begin{figure}
\begin{center}
\includegraphics[width=0.49\linewidth]{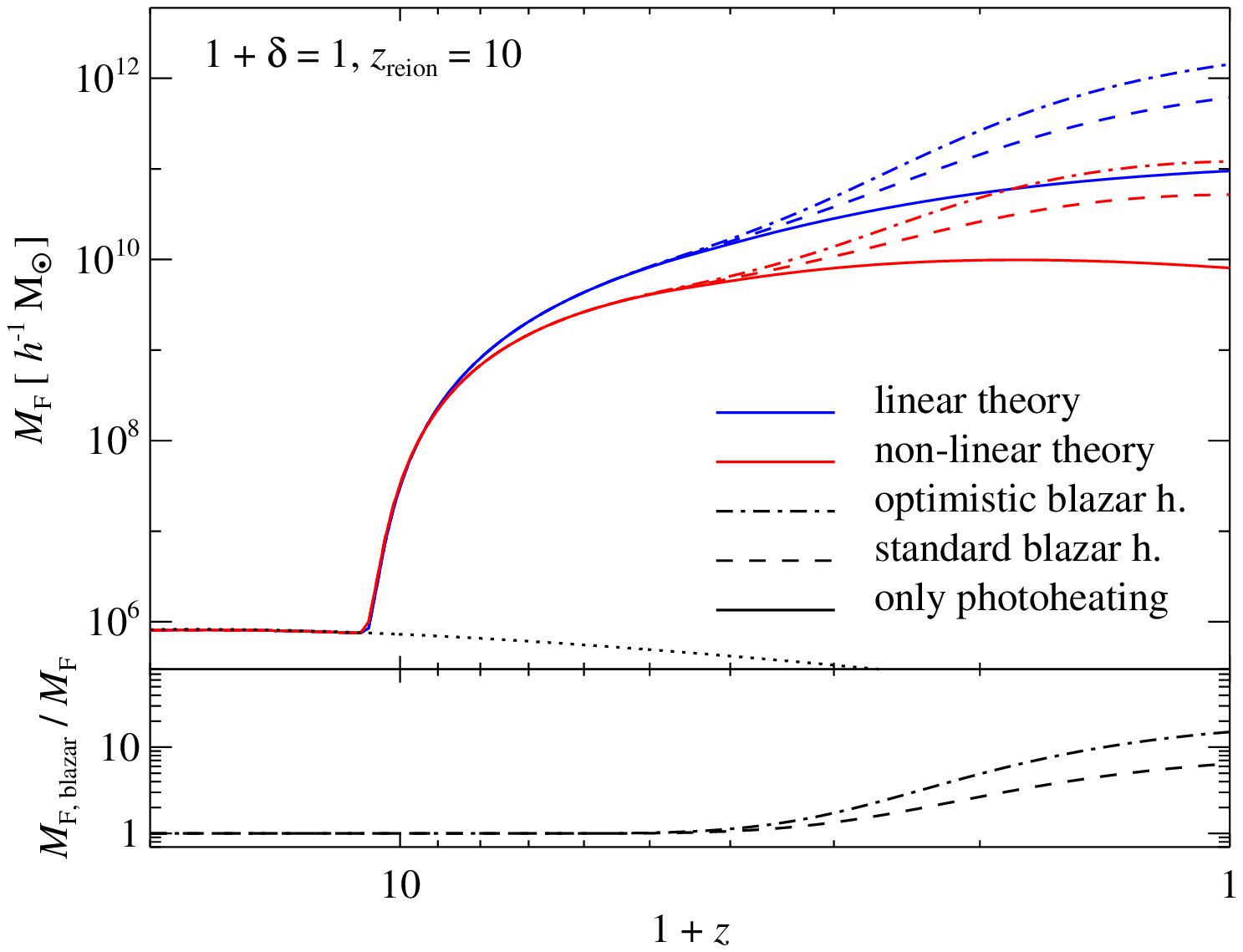}
\includegraphics[width=0.49\linewidth]{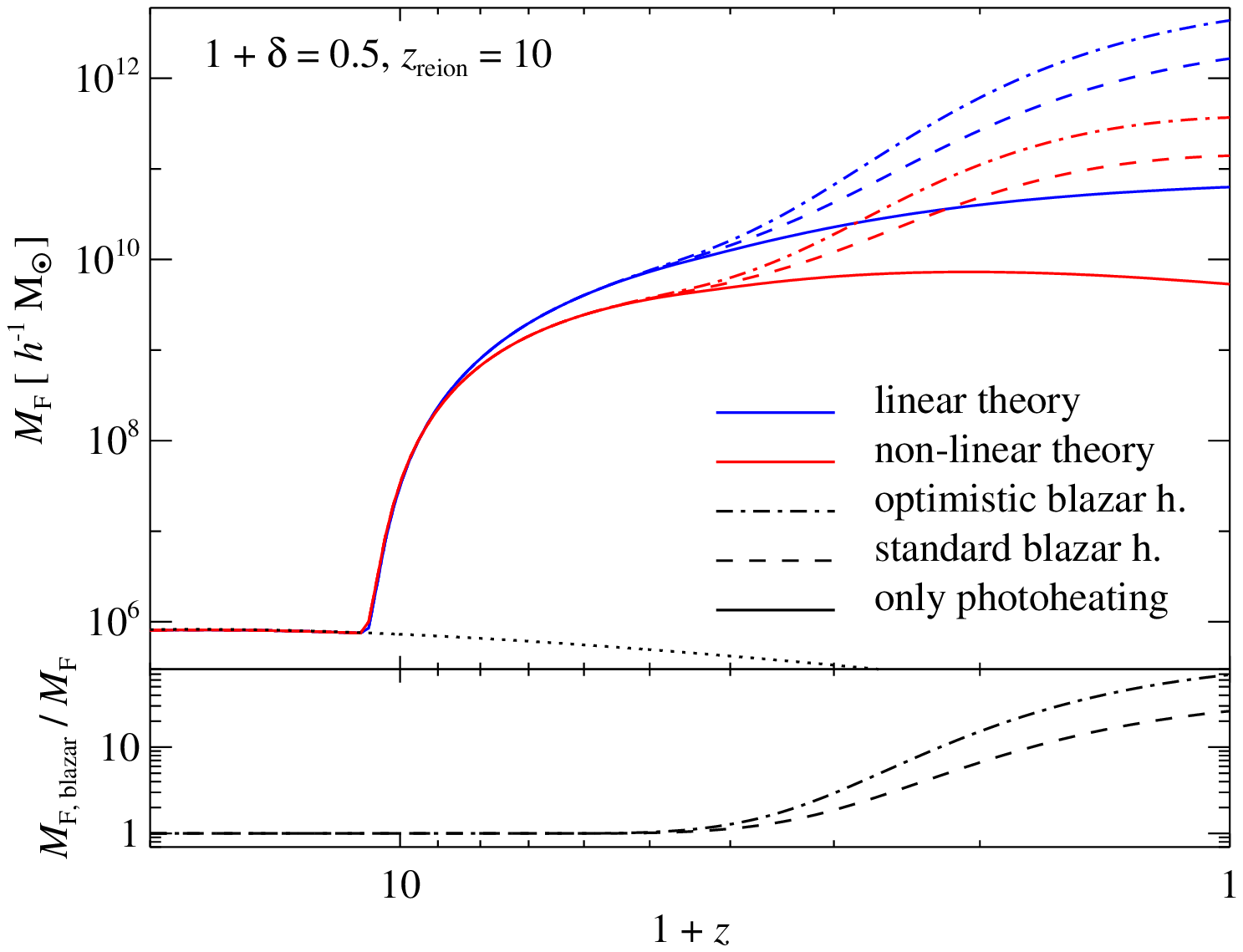}
\end{center}
\caption{Blazar heating suppresses the formation of late-forming dwarf
  galaxies. Redshift evolution of the filtering mass, $M_F$, for the cosmic mean
  density, $\delta=0$ {\it (left)} and for a void with mean overdensity,
  $\delta=-0.5$, {\it (right)}. We contrast $M_F$ in the standard cosmology that
  employs only photoheating (solid) to the case of blazar heating in our
  standard model (dashed) and optimistic model (dash-dotted). In the bottom
  panels, we show the ratio of $M_F$ in our respective blazar heating models to
  those without blazars.  To estimate the effect of nonlinear structure
  formation on the filtering mass, we compare the linear theory $M_F$ (blue) to
  the nonlinear theory $M_F$ (red) where we used a correction function derived
  from hydrodynamic simulations (from Pfrommer {\it et al.} 2012).}
\label{fig:M_f}
\end{figure}

However, there are complications due to non-linear collapse and a delayed
pressure response in an expanding universe. This causes a slight reduction of
the suppression factor (Fig.~\ref{fig:M_f}). Hence, our redshift-dependent
entropy enhancement due to blazar heating increases the characteristic halo mass below
which dwarf galaxies cannot form by a factor of approximately 10 (50) at mean
density (in voids) over that found in the standard model, preventing the
formation of late-forming dwarf galaxies. This may help resolve the ``missing
satellites problem'' in the Milky Way of the low observed abundances of dwarf
satellites compared to cold dark matter simulations and may bring the observed
early star formation histories into agreement with galaxy formation models. At
the same time, it is a very plausible explanation of the ``void phenomenon''
(Peebles \& Nusser 2010) by suppressing the formation of galaxies within
existing dwarf halos, thus reconciling the number of dwarfs in low-density
regions in simulations and the paucity of those in observations.

Finally, this suggests a scenario for the origin of the cool core/non-cool core
bimodality in galaxy clusters and groups, which are separated into different
classes depending on their core temperatures. Early forming galaxy groups are
unaffected because they can efficiently radiate the additional entropy,
developing a cool core. However, late-forming groups do not have sufficient time
to cool before the elevated entropy enhancement is gravitationally reprocessed through
successive mergers---counteracting cooling and potentially raising the core
entropy further to potentially form a non-cool core cluster.

\section{Conclusions and Outlook}

In a series of papers, we have proposed a novel plasma-astrophysical mechanism
that promises transformative and potentially radical changes of our
understanding of gamma-ray astrophysics and the physics of the intergalactic
medium. This can also alter our picture of the formation of dwarf galaxies and
galaxy cluster thermodynamics. Detailed comparisons of predictions of blazar
heating with Lyman-$\alpha$ forest data and \Fermi observation of blazar
statistics as well as the isotropic and anisotropy gamma-ray backgrounds have
been very successful and encouraging.

Nevertheless, we are clearly only beginning to explore the process and
implications of plasma-instability driven blazar heating. Many aspects are only
poorly understood and are now starting to be investigated, including the physics
of the instability in the regime of non-linear saturation. Detailed cosmological
simulations of blazar heating are critical in understanding its impact on
non-linear structure formation. We hope that this work motivates fruitful
observational and theoretical efforts toward consolidating the presented picture
or to modify parts of it.

\section*{Acknowledgments}

C.P.~gratefully acknowledges financial support of the Klaus Tschira Foundation.
A.E.B.~receives financial support from the Perimeter Institute for Theoretical
Physics and the Natural Sciences and Engineering Research Council of Canada
through a Discovery Grant. Research at Perimeter Institute is supported by the
Government of Canada through Industry Canada and by the Province of Ontario
through the Ministry of Research and Innovation.  P.C. gratefully acknowledges
support from the UWM Research Growth Initiative and from \Fermi Cycle 5 through
NASA grant NNX12AP24G. E.P. acknowledges support by the DFG through Transregio
33.

\end{document}